\documentclass{IEEEtran4PSCC}

\usepackage[pdftex]{graphicx}

\usepackage{array}
\usepackage{multirow}
\usepackage{bm}
\usepackage{mdwmath}
\usepackage{xcolor,soul,framed}
\usepackage{cite}
\usepackage{eqparbox}
\usepackage{url}
\usepackage{amsmath}
\usepackage{amssymb}
\usepackage{textcomp}
\usepackage{bm}
\usepackage{dblfloatfix}
\newcommand{\busesm}{\mathcal{N}}

\newcommand{\meas}{$\mathcal{M}$}
\newcommand{\measm}{\mathcal{M}}
\newcommand{\pseudom}{$\mathcal{P}$}
\newcommand{\mpseudom}{\mathcal{P}}

\usepackage[caption=false, font=normalsize, labelfont=sf, textfont=sf]{subfig}

\bstctlcite{IEEEexample:BSTcontrol}

\begin{document}

\title{Exact Modeling of Non-Gaussian Measurement Uncertainty in Distribution System State Estimation}

\author{Marta~Vanin,~\IEEEmembership{Member,~IEEE, }
 Tom~Van~Acker,	Reinhilde~D'hulst, 
       and~Dirk~Van~Hertem,~\IEEEmembership{Senior Member,~IEEE}

\thanks{M. Vanin, T. Van Acker, and D. Van Hertem are with the Research Group ELECTA, Department of Electrical Engineering, KU Leuven, 3001 Heverlee, Belgium. 
Reinhilde D'hulst is with VITO, Boeretang 200, 3400 Mol, Belgium.
M. Vanin, R. D'hulst, and D. Van Hertem are also with EnergyVille, Thor Park 8310, 3600 Genk, Belgium.
 Corresponding author: marta.vanin@kuleuven.be
 }
\thanks{This work is supported by the  Flemish DSO Fluvius within the framework of the research project ADriaN – \textit{Active Distribution Networks.}}}

\maketitle

\begin{abstract}
State estimation allows to monitor power networks, exploiting field measurements to derive the most likely grid state. In the literature, measurement errors are usually assumed to follow zero-mean Gaussian distributions; however, it has been shown that this assumption often does not hold. One such example is when considering pseudo-measurements. In distribution networks, a significant amount of pseudo-measurements might be necessary, due to the scarcity of real-time measurements. In this paper, a state estimator is presented which allows to model measurement uncertainty with any continuous distribution, without approximations. This becomes possible by writing state estimation as a general maximum-likelihood estimation-based constrained optimization problem. To realistically describe distribution networks, three-phase unbalanced power flow equations are used. Results are presented that illustrate the differences in accuracy and computational effort between different uncertainty modeling methods, for the IEEE European Low Voltage Test Feeder. 
\end{abstract}


\section{Introduction}\label{sec:introduction}
\subsection{Background and Motivation}

\IEEEPARstart{T}{he} ongoing electrification of transport and heating systems, as well as the installation of distributed generation, are changing the role of the distribution system, calling for more active management strategies~\cite{ANM}. In this light, state estimation (SE) gives the potential to better understand the distribution system (DS) and serves as a basis to successively perform optimization and control actions~\cite{towardsSG}. While SE is industrial practice in the transmission system, the same SE techniques cannot be directly applied to the DS, as this inherently presents different features and additional implementation challenges, which have been discussed in numerous papers, e.g.,~\cite{towardsSG, dellagiustina, Vanin2022}. One of the challenges towards the design and implementation of reliable Distribution System SE (DSSE) is the paucity of available measurement devices, which results in the need for forecasts of non-metered loads and generators to obtain full system observability. These forecasted values are called \textit{pseudo-measurements}, and their accuracy in representing the actual components has a significant impact on the SE quality~\cite{Angioni}. The standard assumption in the SE practice is that all measurements and pseudo-measurements are characterized by zero-mean Gaussian error distributions~\cite{bible}, which allows to employ well-known SE techniques, such as weighted least squares (WLS) SE. However, it has long been known that residential demand and renewable generation forecasts are better described with non-Gaussian distributions, e.g., Beta or LogNormal~\cite{Heunis, Ghosh, Bludszuweit}, or with Gaussian mixture models (GMM), which are relatively popular for DS demand modelling~\cite{SinghGMM}. 

To be compatible with standard WLS SE calculations, these non-Gaussian uncertainties ultimately have to be approximated to one equivalent Gaussian probability distribution function (pdf), which results in approximations in the uncertainty models. Thus, methods have been pursued that allow to include better statistical models within the estimators~\cite{McLoughlin2015}. However, these often require ``non-standard" re-formulations of the SE problem.

\subsection{Related Work}

Recent analyses on field data revealed that zero-mean Gaussian assumptions do not hold for phasor measurement units (PMUs)~\cite{Wang_PMU}, raising interest in better error representations for transmission system SE, where these devices are commonly present. For this purpose, a dynamic generalized maximum-likelihood unscented Kalman filter was devised~\cite{Zhao_GMUKF}. While dynamic SE has clear benefits and implementation possibilities in the transmission system, a number of publications observed that dynamic DSSE does not improve on static WLS~\cite{Huang}, unless a large number of PMUs is installed~\cite{Carquex,Song}, which is unrealistic in a DS context. Thus, the focus of this paper is on developing a method for static DSSE, in which the smart meters (SM) and pseudo-measurements are the only available measurement inputs to SE, which is a realistic DS setup. 

The PMU error considerations in~\cite{Wang_PMU} motivated work on maximum-likelihood estimation (MLE) for static SE~\cite{Chen2021,Chen2020}, again in a transmission context. While~\cite{Chen2020} seems the first realistic application of MLE for balanced SE, an earlier similar conceptualization can be found in~\cite{Minguez}. The present work also relies on MLE, but addresses distribution systems, utilizing a novel optimization-based approach. An interesting non-Gaussian DSSE was first proposed in~\cite{Angioni2016} and improved in~\cite{PegoraroBayesian}, in which statistical information is included with a Bayesian approach. However, the method proved computationally intensive and was only tested on a small balanced test case. In~\cite{Brueggerman2017}, particle filter is applied to DSSE. This is also a Bayesian type of filter, but it approximates the probability distributions by sampling them instead. While these approaches allow to represent less usual and even discrete distributions, they are significantly more expensive from a computational standpoint, and their published applications to power system SE to date involve relatively small-sized systems~\cite{Brueggerman2017,Alhalali2015,Cevallos2018}. Furthermore, modelling the particles/samples evolution in the state-space over time is required, which is challenging in distribution networks where load variations are non-smooth~\cite{Alhalali2015}.

\subsection{Contributions and Outline of the Paper}

To the best of the authors' knowledge, and as per recent SE review papers~\cite{Primadianto, Dehghanpour}, the remainder of the DSSE literature resorts to zero-mean Gaussian assumptions or approximations, and this is the first work on MLE-based SE with unbalanced power flow equations, which are required for a realistic representation of DS~\cite{DSSE95}. This work further differentiates itself from transmission MLE papers~\cite{Chen2021, Chen2020} by formulating DSSE as a constrained non-convex problem, based on a generic ``optimization-first" DSSE set-up like that in~\cite{Vanin2022}. From a MLE-SE standpoint, this allows to use any continuous pdf to describe (pseudo-)measurements, as well as having a combined objective to separately consider Gaussian and non-Gaussian quantities, exploiting the computational advantages of Gaussian assumptions when possible. 
Finally, it should be noted that many common pdfs are log-concave~\cite{Bagnoli}, which would allow the use of convex optimization techniques, if the underlying power flow formulation is also convex.

To summarize, the main contribution of this work is the formulation of a novel and scalable DSSE, that allows the exact modeling of smooth non-Gaussian uncertainty in unbalanced networks. This is made possible through a generalized constrained-optimization reformulation of DSSE that allows to realize a wide variety of different MLE-SE scenarios. Given the modular and extendable problem formulation, (in)equality constraints can easily be added to address specific issues, e.g., enforcing fixed power factor. In this work, specific constraints have been introduced to address rooftop PV generation and reactive power models. These add correlations between power generation variables, which is also a variation on standard DSSE, in which all random variables are usually assumed to be independently distributed~\cite{bible}. The MLE-SE is based on the DSSE concepts described in~\cite{Vanin2022}, and is available open-source\footnote{https://github.com/Electa-Git/PowerModelsDistributionStateEstimation.jl}. It is possible and relatively easy for the readers/users to further extend the code to tackle additional MLE-SE cases. Finally, this paper illustrates various ways to model measurement uncertainty in DSSE, and can be used as a tutorial for this purpose.

The rest of the paper is structured as follows: Section~\ref{sec:mathematical_model} presents the mathematical model of the optimization-based DSSE, Section~\ref{sec:case_studies} describes the uncertainty models used in this work, and Section~\ref{sec:results} shows numerical results for a low voltage feeder. Finally, conclusions are drawn in Section~\ref{sec:conclusions}.
\section{Mathematical Model and Implementation}\label{sec:mathematical_model}
\newcommand{\z}{$\mathbf{z}$}
\newcommand{\h}{$\mathbf{h}$}
\newcommand{\hhm}{$\mathbf{h}_m$}
\newcommand{\errv}{$\boldsymbol{\eta}$}
\newcommand{\vm}{$|U|$}
\newcommand{\va}{$\angle U$}
\newcommand{\vmjp}{$|U_{j,p}|$}
\newcommand{\vajp}{$\angle U_{j,p}$}
\newcommand{\vi}{$U^{\text{im}}$}
\newcommand{\vr}{$U^{\text{re}}$}
\newcommand{\ca}{$\angle I $}
\newcommand{\cax}{$\angle I_c $}
\newcommand{\cix}{$I^{\text{im}}_c$}
\newcommand{\crx}{$I^{\text{re}}_c$}
\newcommand{\cmx}{$|I|_c$}
\newcommand{\px}{$P_c$}
\newcommand{\qx}{$Q_c$}
\newcommand{\w}{$W$}

\newcommand{\cixm}{I^{\text{im}}_c}
\newcommand{\crxm}{I^{\text{re}}_c}
\newcommand{\cmxm}{|I|_c}
\newcommand{\vim}{U^{\text{im}}}
\newcommand{\vrm}{U^{\text{re}}}
\newcommand{\cam}{\angle I}
\newcommand{\caxm}{\angle I_c}
\newcommand{\vmm}{|U|}
\newcommand{\vam}{\angle U}
\newcommand{\pxm}{P_c}
\newcommand{\qxm}{Q_c}
\newcommand{\wm}{W}

\newcommand{\N}{\textbf{\textcolor{blue}{N}}}

\newcommand{\varspace}{$\mathcal{X}$}
\newcommand{\varspacem}{\mathcal{X}}

\subsection{MLE-SE Model}

SE derives the most-likely state of a network based on a set of measurements \meas \ and pseudo-measurements \pseudom. This consists of assigning numerical values to the network’s variable space $\varspacem^{\text{form}}$, which is defined by the chosen power flow formulation. The general problem structure of the proposed DSSE approach is similar to that of an optimal power flow (OPF) problem: 
\begin{IEEEeqnarray}{ l C l }
    \text{minimize} \; \;       &~& \sum_{\substack{j \in \measm \cup \mpseudom }}   \rho_j,         \label{eq:objective}    \\
    \text{subject to:}          &~& \mathbf{f}(\boldsymbol{\rho}, \mathbf{x}) = 0,   \label{eq:f}            \\
                                &~& \mathbf{h}(\mathbf{x}) = 0,                      \label{eq:h}            \\    
                                &~&  \mathbf{k}(\mathbf{x}) \leq 0, \label{eq:k}
\end{IEEEeqnarray}
where \textbf{x} is the state vector and $\rho_j$ are the real- and pseudo-measurement residuals. Eq. \eqref{eq:h} represents the power flow equations, in the form of equality constraints, while \eqref{eq:k} describes any inequality constraint of the problem. These can be, for instance, bounds on the variables that limit the search space to achieve faster convergence. In the present work, an ``AC" rectangular (ACR) power flow formulation is used, so the variable space consists of power and voltage in rectangular coordinates: $\mathbf{x} \in \varspacem^{\text{ACR}} = \{ \vrm, \vim, P, Q \}$, and \eqref{eq:h} can be compactly summarized as follows:
\begin{align}
& \mathbf{S}_{ij} = \mathbf{U}_i \mathbf{U}_i^\text{H} (\mathbf{Y}_{ij}+\mathbf{Y}^c_{ij})^\text{H} - \mathbf{U}_i \mathbf{U}_j^\text{H} \mathbf{Y}_{ij}^\text{H}, \; \; \; \forall (i,j) \in \mathcal{E} \label{eq:SijBIM}, \\
& \mathbf{S}_{ji} = \mathbf{U}_j \mathbf{U}_j^\text{H} (\mathbf{Y}_{ij}+\mathbf{Y}^c_{ji})^\text{H} - \mathbf{U}_i^\text{H} \mathbf{U}_j \mathbf{Y}_{ij}^\text{H}, \; \; \;  \forall (i,j) \in \mathcal{E} \label{eq:SjiBIM}, \\
& \mkern20mu \sum_{k \in G_i} \mathbf{S}_k^g - \sum_{k \in L_i} \mathbf{S}^d_k = \mkern-20mu \sum_{(i,j) \in \mathcal{E}_i \cup \mathcal{E}_ i^R} \mkern-20mu \text{diag}(\mathbf{S}_{ij}), \; \; \forall i \in \busesm,
\end{align}
where $\mathbf{U}_i$ is the 3$\times$1 vector of the voltage phasors for bus $i$, $\mathbf{S}_{ij}$ is the 3$\times$3 matrix of the apparent power flow in branch $(i,j)$, $\mathbf{S}^g_{k},\mathbf{S}^d_{k}$ are the generation and demand from device $k$, $\mathcal{N}$ is the set of feeder buses, $\mathcal{E}$ and $\mathcal{E}^{R}$ are the sets of branches in the forward and reverse orientation, $L_i$ and $G_i$ are the sets of loads and generators at bus $i$. Finally, $\mathbf{Y}_{ij}, \mathbf{Y}^c_{ij}$ are the series and shunt admittance of the 3$\times$3 $\pi$-model used to model branches, as illustrated in~\cite{PMD_PSCC}. The voltage angle of a reference bus is assigned, as is common SE practice \cite{bible}:
\begin{equation}\label{eq:refangle}
\angle \mathbf{U}^{ref} = [0, -2/3\pi, 2/3\pi].     
\end{equation}
It can be observed that the use of \eqref{eq:h} allows to include zero-injection buses as equality constraints, avoiding the need for ill-conditioning weighted virtual measurements~\cite{Clements}. Finally, \eqref{eq:f} is the residual definition, namely the function that associates the (pseudo)-measurements to the system variables. If all variables~$x_{j}$ can be described by univariate continuous distributions with associated pdf:~$f_j(x_{j})$, the most-likely state of the system can be found through MLE \cite{Minguez}: 
\begin{equation}\label{eq:MaxProduct}
    \text{maximize} \prod\limits_{j \in \measm \cup \mpseudom} f_{j}(x_j).
\end{equation}

To get rid of the exponential terms of common pdfs, it is equivalent and more convenient to maximize the log-likelihood:

\begin{equation}\label{eq:logMLE}
    \text{maximize} \sum\limits_{j \in \measm \cup \mpseudom} \log (f_{j}(x_j)).
\end{equation}

Note that if all $x_j$ follow a Gaussian distribution, \eqref{eq:logMLE} is equivalent to the standard WLS SE\cite{Minguez}:
\begin{equation}\label{eq:WLS}
    \text{minimize} \sum_{j \in \measm \cup \mpseudom} \frac{(x_j-\mu_j)^2}{\sigma_j^2},
\end{equation}
and to a weighted least absolute value~\cite{Vanin2022} minimization if $x_j$ is Laplacian. In \eqref{eq:WLS}, $\mu_j$ indicates the (pseudo)-measurement value, and $1/\sigma_j^2$ its weight, which is the inverse of the distribution variance and depends on the measurement accuracy. From a computational standpoint, it is more convenient to use the simplified notation in \eqref{eq:WLS} rather than \eqref{eq:logMLE}. The same holds if the distribution is assumed Laplacian, in which case the residuals can be formulated as linear inequality constraint, as illustrated in~\cite{Vanin2022}, with no computational burden. Assuming that SM errors are Gaussian distributed, while pseudo-measurements are not, \eqref{eq:f} becomes:
\begin{align}\label{eq:rho_m_rho_p}
   \rho_p = \xi_p - \log(f_p(x_{p})),  \; \; \; \; \; \; \; \; \forall p \in \mpseudom,\\    
  \rho_m = (x_m - z_m)^2/\sigma^2_m,  \; \; \; \; \; \forall m \in \measm,   
\end{align}
where $z_m \in \mathbb{R}$ is the measured value and $\xi_p$ shifts the function to ensure that $\rho_j \geq 0, \; \; \forall j \in \measm \cup \mpseudom$:
\begin{equation}\label{eq:shift}
\xi_p = |\text{minimum}\{ -\log(f_p(x_{p})) \}|, \; \; \forall p \in \mpseudom.    
\end{equation}

Finally, the minus sign before $\log(f_p(x_{p}))$ in \eqref{eq:rho_m_rho_p} allows to combine the different residual definitions in a single objective \eqref{eq:objective}. Note that the assumption that SM errors are zero-mean Gaussian, while pseudo-measurements are not, is realistic but arbitrary. The assumption is made in this paper for illustrative purposes, to: $1)$ show that the proposed method can combine Gaussian and non-Gaussian measurement sources, and $2)$ allow to analyze the computational efforts for increasing amounts of non-Gaussian errors.

\subsection{MLE-SE Implementation}

The MLE-SE is made available as part of a Julia package called PowerModelsDistributionStateEstimation.jl, which is an extension of PowerModelsDistribution.jl~\cite{PMD_PSCC}, and based on the mathematical programming toolbox JuMP~\cite{JuMP}. It also makes substantial use of Distributions.jl~\cite{DistributionsJL}. The implementation is such that modeling and solving layer are separate, so that users can focus on designing a suitable DSSE formulation and delegate the solution to an off-the-shelf solver. This allows to rapidly experiment with, e.g., different SE formulations or measurement uncertainty models, and makes the tool ``topology agnostic": it works for meshed and radial feeders alike. PowerModelsDistributionStateEstimation.jl is designed to quickly develop and test different SE methods, rather than developing the computationally most performant implementation. As such, once a suitable SE model is found, users might choose to develop a customized implementation to enhance computational efficiency. Nevertheless, this paper's results show that the package's MLE-SE computational times seem already acceptable for static DSSE.

The tool currently supports the following distributions out of the box: \textit{Laplacian}, \textit{Beta}, \textit{Gamma}, \textit{Weibull}, \textit{Gaussian}, \textit{LogNormal}, \textit{GMM}, and $Polynomial$ fits. While these seem to encompass most realistic cases, it is also straightforward to extend the tool to use any distribution, as long as its pdf (``$f$") or log-pdf (``$\log(f)$") is smooth and twice differentiable. 
\section{Case Studies}\label{sec:case_studies}

The case studies illustrate the trade-off between SE accuracy and computational effort for each uncertainty model, depending on the amount of pseudo-measurements in the feeder. In a real-life setting, this type of analysis can be performed a priori, to decide which uncertainty model to use for operations. Once the optimal model is chosen, the tool/method can be used for routine SE calculations. 

\subsection{Model of Pseudo-measurement Uncertainty}\label{sec:uncertainty}

In this work, four distributions are considered, amongst those supported out-of-the-box in the tool: $Beta$ and $Polynomial$ are considered to be the two ``original" distributions, i.e., those that exactly describe the power variable, and are used in the case studies alongside three of their possible approximations, two of which are $Gaussian$, and one is a $GMM$. Historical power demand data for Belgian residential consumers show that $Beta$ distributions make for realistic forecast models. The $Polynomial$ case has been chosen as it is a generic and flexible way to represent variables that do not follow a conventional pdf.  

\subsubsection{Beta Distribution} the pdf reads:
\begin{equation}\label{eq:beta_pdf}
    f(x) = \frac{(x-x^{min})^{\alpha-1}(x^{max}-x)^{\beta-1}}{B(\alpha, \beta)(x^{max}-x^{min})^{\alpha+\beta-1}},
\end{equation}
where $B(\alpha, \beta)$ is the Beta function:
\begin{equation}
    B(\alpha, \beta) = \int_0^1 x^{\alpha-1}(1-x)^{\beta-1}dx.
\end{equation}

\subsubsection{Polynomial}
This can be a generic polynomial of any degree, where the degree is chosen in a curve-fitting pre-processing step. To reduce complexity, it is more practical to derive the polynomial fit of the log-pdf instead of the pdf function, to avoid the logarithms in \eqref{eq:logMLE}. This approach is adopted in the present exercise.

\subsubsection{Gaussian Mixture Model}

GMM consists of approximating a generic probability density function as the weighted sum of a finite number ($G_c$) of Gaussian components' pdfs:
\begin{equation}\label{eq:GMM}
    f(x) \approx f^{\text{GM}}(x) = \sum_{i=1}^{G_c} \omega_i \frac{e^{-0.5\cdot (\frac{x-\mu_i}{\sigma_i})^2}}{\sigma_i \sqrt{2\pi}},
\end{equation}
with the condition that the sum of the weights is unitary:
\begin{equation}
    \sum_{i=1}^{G_c} \omega_i = 1.
\end{equation}

\subsubsection{Gaussian Equivalent (GE)}

To fit a standard WLS approach, $f^{\text{GM}}(x)$ needs to be converted into an equivalent Gaussian distribution $f^{\text{GE}}(x)$ \cite{SinghGMM}, with parameters $\sigma^{eq},\mu^{eq}$, thus performing a further approximation of the original distribution. To do so, a subset of the $G_c$ Gaussian components of the GMM is chosen. Let $E_c$ be these components. There are several approaches to decide which and how many of the GMM components to keep to derive the GE model; a discussion can be found in \cite{Crouse}. Once the $E_c$ are obtained, the GE distribution is defined by its $\mu^{eq}$ and $\sigma^{eq}$~\cite{SinghGMM}:
\begin{align}
& \omega^{eq} = \sum_{j \in E_c} \omega_j, \\
& \mu^{eq} = \frac{1}{\omega_m} \sum_{j \in E_c} \omega_j \mu_j,  \\
& (\sigma^{eq})^2 = \frac{1}{\omega^{eq}} \sum_{j \in E_c} \omega_j \cdot ( \sigma_j^2 + (\mu_j - \mu^{eq})^2).    
\end{align}

\subsubsection{Gaussian Approximation (GA)} \label{sec:ga}

This is the simplest approximation and is a Gaussian fit of the original distribution.

The GMM and GA are fitted from the original distribution via MLE, which is conceptually similar to the MLE of the presented DSSE, but has the aim of finding the GMM and GA parameters, and is common practice in statistics \cite{CousineauMLE}. Since they are \textit{Gaussian} distributions, when GA or GE models are used for $p \in \mpseudom$, $\rho_p = (x - \mu_p)^2\cdot(\sigma_p)^{-2}$. In the case studies, measurement uncertainty for users with a SM is also assumed to be \textit{Gaussian}. This means that the GE and GA cases can be effectively formulated as WLS SE, which solves faster than the MLE SE. Including GMM, or any other polynomial or parametric distribution in its full form, on the other hand, requires the more general MLE-SE (note again that WLS-SE is also a specific case of MLE-SE).

\subsection{Pseudo-measurements for PV injection}

Pseudo-measurements are forecasted values of consumer demand/generation, and as such the $x$ in \eqref{eq:beta_pdf}-\eqref{eq:GMM} can represent active/reactive power, solar irradiance, or other desired quantities in general. With the mathematical model discussed so far, it is possible (although not mandatory) to assign different pdfs to every user. The realizations of the pdfs are independent, and as such the individual power injections $x_i$ are not correlated. Non-correlation is a standard assumption in SE~\cite{bible}, and while it is typically considered acceptable for power demand, PV power injections in a feeder are strongly correlated: the irradiance in a restricted geographical area is roughly the same. Thanks to the flexibile optimization-first formulation, to include this in the DSSE problem, it is sufficient to add the following constraint: 
\begin{equation}\label{eq:pv}
    x_i = x_j, \;\; \; \forall \, (i,j) \; \in \; \mpseudom.
\end{equation}
The above could be applied to a subset of $\mpseudom$ only, if pseudo-measurements include both (non-correlated) non-monitored demand and (correlated) non-monitored PV generation. In this paper, we examine PV-only cases. 

Furthermore, we assume for simplicity that all non-monitored users have the same installed PV capacity. When this is not the case, to convert the irradiances $x_i$ in \eqref{eq:pv} into power, it is sufficient to multiply $x_i$ according to the PV installation size of user $i$. 

\subsection{Assumptions on Reactive Power}
In this paper, we consider two different ways to model reactive power: $1)$ for every user, Q is independent from P, and has its own distribution, $2)$ for every user, the power factor is constant. The latter is a common assumption in distribution system modeling, and can realistically take place if inverter-connected devices adopt this type of control setting.
To implement model $1)$, in this paper we assume that reactive power pseudo-measurements pdfs are simply rescaled active power pdfs: 
\begin{equation}\label{eq:reactive_independent}
 f^Q(x)~=~k_1 \cdot f^P(x).    
\end{equation}
This is again an arbitrary choice. Analyzing the impact of different pdfs for reactive power is out of the scope of this paper.
To enforce constant power factor, on the other hand, it is sufficient to not include Q measurements in $\mpseudom$ and add the following constraint to the existing model:
\begin{equation}\label{eq:const_pf}
    Q_i = k_2 \cdot P_i \; \; \forall \;i \;\; \text{s.t.} \;\; P_i \in \mpseudom.
\end{equation}
In this paper, we assume $\cos \phi = 0.95$ for all users, and $k_i$ in both \eqref{eq:reactive_independent} and \eqref{eq:const_pf} is $\tan \phi$ for simplicity.
The constant power factor approach reduces the number of (non-Gaussian, in this case) measurement residuals. Nothing prevents the adoption of a hybrid approach where some of the users are modelled with $1)$ and some with $2)$.

\section{Numerical Results}\label{sec:results}
\newcommand{\vmmbetai}{|U_i^{\text{\Beta}}|}
\newcommand{\vmmsei}{|U_i^{\text{se}}|}

\begin{figure*}[t]
\centering
\begin{tabular}{cc}
  \includegraphics[width=0.45\textwidth]{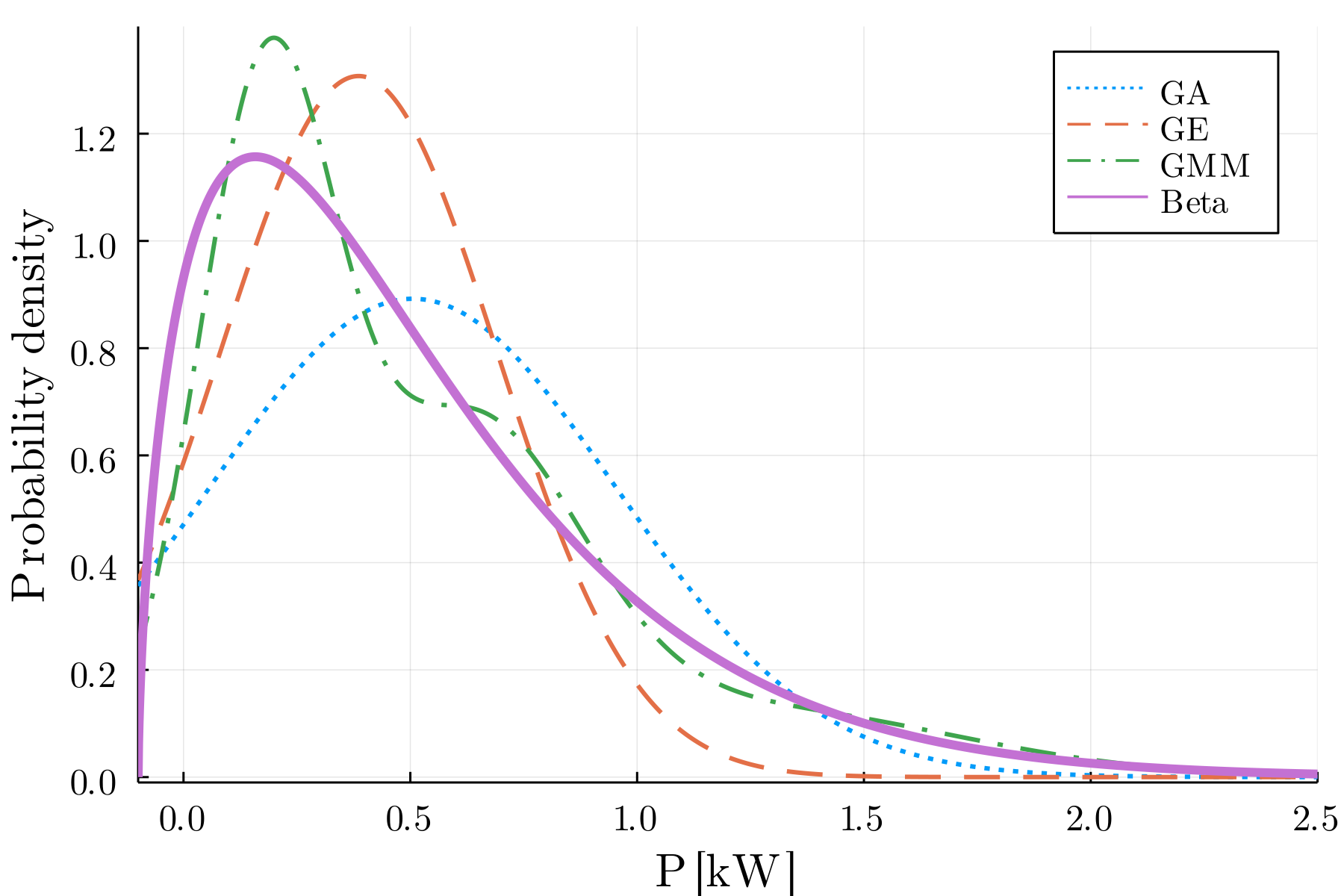}\label{fig:beta_pdf} &   \includegraphics[width=0.45\textwidth]{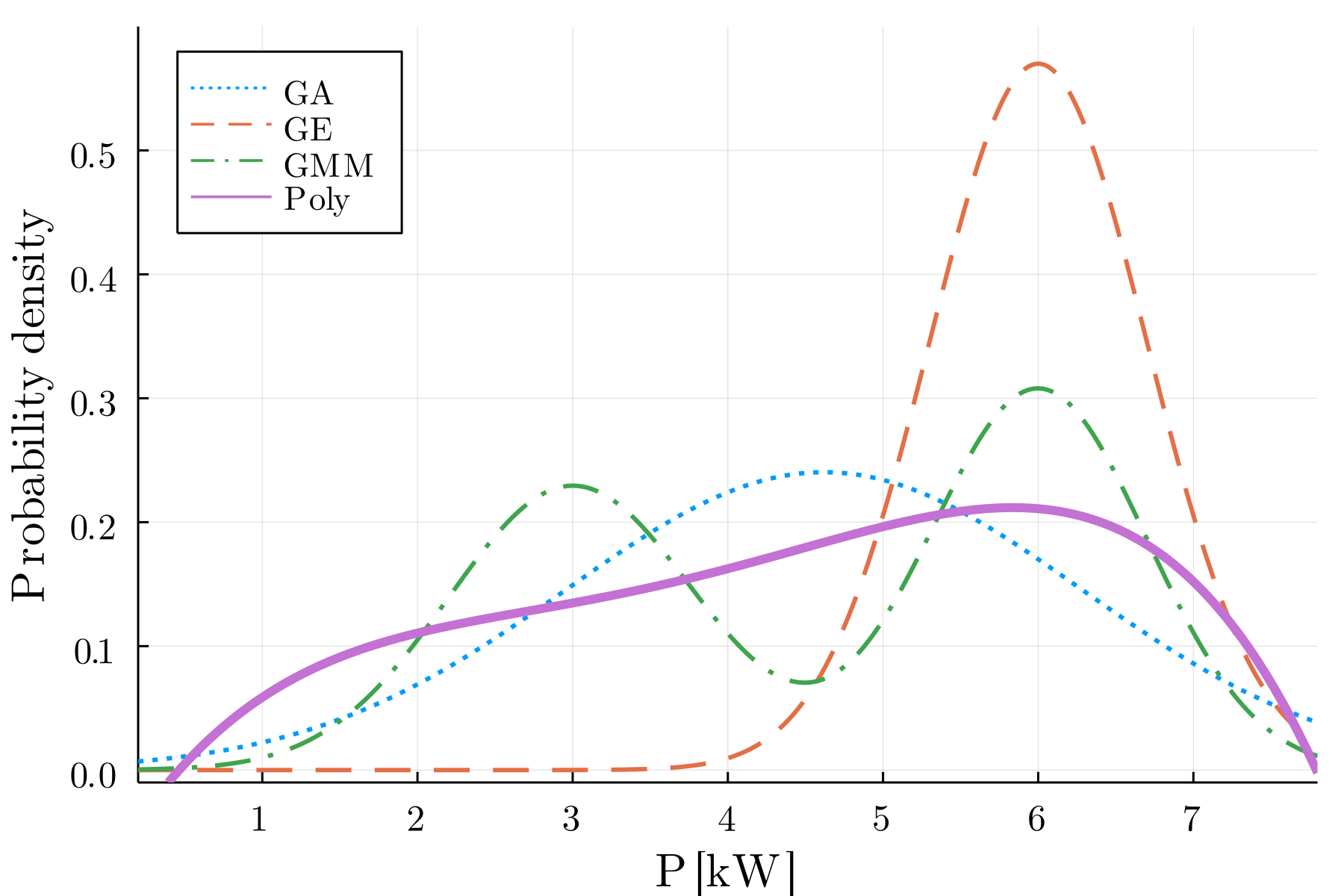}\label{fig:poly_pdf} \\
\end{tabular}
\caption{Pdfs of the chosen Beta and Polynomial distributions and their approximations.}
\label{fig:pdfs}
\end{figure*}

The parameters of the chosen distributions are reported in Tables~\ref{tab:parameters_set_1}-\ref{tab:parameters_set_2}, and their pdfs are shown in Fig.~\ref{fig:pdfs}. In the case studies, the $Beta$ and $Polynomial$ distributions are assumed to be the best forecast representation, while GA, GE and GMM are all approximations. This is only an illustrative example, and not necessarily always true. In particular, GMM are often used for power demand, and could be the ``reference" pdf in a different test case. 

To assess the trade-off between the accuracy and solve time of the different SE models, power flow (PF) and SE calculations are performed on the IEEE Low Voltage Test Feeder~\cite{EU_LV}. The feeder originally presents 55 users and 906 buses, which have been reduced to 118 (without approximations), as illustrated in~\cite{ClaeysCIRED2021}. 
Power flow results are indicated with a $^{PF}$ superscript, SE results, with $^{SE}$.

\begin{table}[b]
\centering
\caption{Parameters for $Beta$ distr. and approximations}
\begin{tabular}{|l|l|}
 \hline
Beta  & $\alpha$ = 1.6339, $\beta$= 20.9022, $x^{\text{min}}$ = -0.1, $x^{\text{max}}$ = 8.268 \\ \hline
GMM  & $\mu$ = [0.181, 1.223, 0.627], $\sigma$ = [0.152, 0.467, 0.241], \\ & $\omega$ = [0.476, 0.152, 0.372]   \\ \hline
GE  &  $\mu$ = 0.386, $\sigma$ = 0.305 \\ \hline
GA    & $\mu$ = 0.505, $\sigma$ = 0.447 \\ \hline
\end{tabular}\label{tab:parameters_set_1}
\end{table}

\begin{table}[b!]
\centering
\caption{Parameters for $Polynomial$ distr. and approximations}
\begin{tabular}{|l|l|}
 \hline
Poly  & -0.080 + 0.209$x$ - 0.086$x^2$ + 0.017$x^3$ - 0.001$x^4$ \\ \hline
GMM  & $\mu$ = [3.0, 6.0], $\sigma$ = [0.80, 0.70], \\ & $\omega$ = [0.46, 0.54]   \\ \hline
GE  &  $\mu$ = 6.0, $\sigma$ = 0.70\\ \hline
GA   & $\mu$ = 4.62, $\sigma$ = 1.66 \\ \hline
\end{tabular}\label{tab:parameters_set_2}
\end{table}

\begin{figure*}[t!] \centering
  \includegraphics[width=0.7\textwidth]{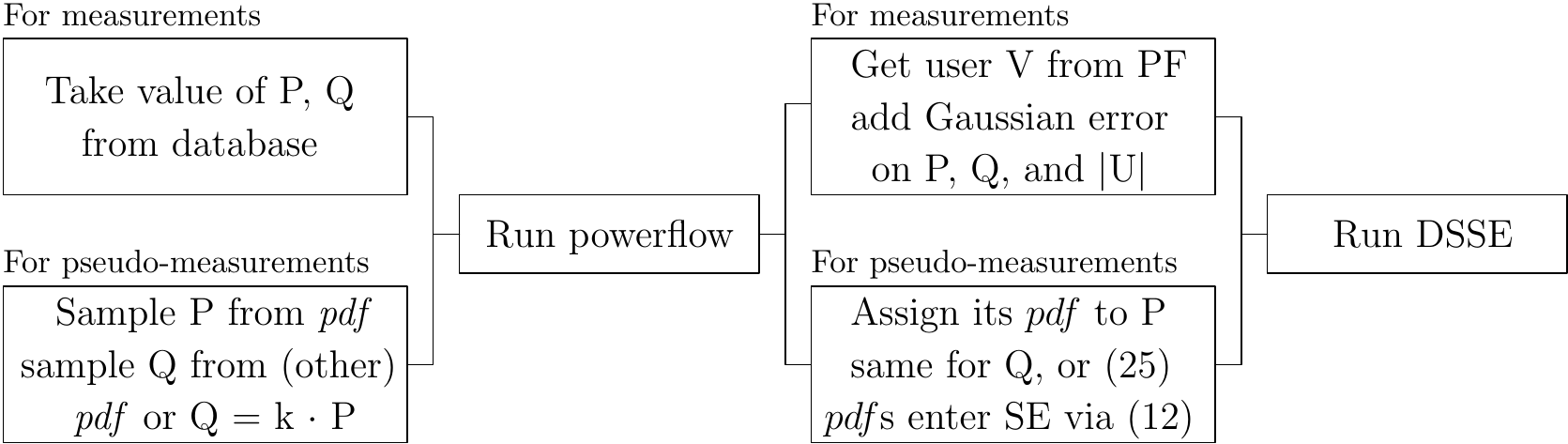}
\caption{To create the SE input, a PF is first run. For users with a SM, the input to the PF are demand values taken from the publicly available IEEE test-case data. For pseudo-measured users, it is a sample from the demand pdfs. For users with SM, an error is added to the original SM demand and to the voltage at the user bus returned by the power flow. For pseudo-measurements, the original demand pdfs is used. }
\label{fig:error_addition}
\end{figure*}

It is assumed that the only available SM measurements are voltage magnitude $\vmm$ and active and reactive power, which is realistic for low voltage networks~\cite{Vanin2022}. It should be noted that the open-source tool supports additional measurement types and power flow formulations. The ACR formulations has been chosen because it is exact and delivers relatively fast DSSE solutions~\cite{Vanin2022}, and P and Q are in its variable space. 
To integrate \vm \, measurements in the optimization problem, since these are not native to $ \varspacem^{\text{ACR}} $, constraint \eqref{eq:vm} is added to \eqref{eq:objective}-\eqref{eq:k}: 
\begin{equation}\label{eq:vm}
\vmm^2 = (\vrm)^2 + (\vim)^2.
\end{equation}

The process to create measurement inputs for the SE is shown in Fig.~\ref{fig:error_addition}.  
 For the voltage measurements, the standard deviation is 0.38~V, which is in line with the accuracy of current SMs. For SM power measurements, $\sigma$ is set to $1/100$ of the $\sigma$ of the GA, to make sure that actual measurements have a higher weight in the overall SE than pseudo-measurements. The $\sigma$ values can affect the SE results and depend on the SM accuracy and on design choices when SM and pseudo-measurements are combined. 

To illustrate the accuracy of the different uncertainty models, the following metrics are used: average ($\Delta U^{\text{avg}}$) and maximum ($\Delta U^{\text{max}}$) voltage magnitude difference, in per unit, and transformer active power difference ($\Delta P^{\text{t}}$), in kW:
\begin{align}
    \Delta U^{\text{avg}} = & \mkern-10mu \sum_{i \in N, \phi \in \Phi_i} ( |\, \vmm^{\text{PF}}_{i,\phi}-\vmm^{\text{SE}}_{i,\phi} \,|)/|N|, \\
    \Delta U^{\text{max}} = & \; \max \{ | \, \vmm^{\text{PF}}_{i,\phi}-\vmm^{\text{SE}}_{i,\phi} \,| \}, \; \; \forall i \in N, \phi \in \Phi_i, \\
    \Delta P^{\text{t}}_{\phi} = & \; |P^{\text{t,PF}}_{\phi}-P^{\text{t, SE}}_{\phi}|, \; \; \forall \phi \in \{a, b, c\},
\end{align}
where $N$ is the set of the feeder buses, and $\Phi_i \subseteq \{a,b,c\}$ the set of the phases to which bus $i$ is connected. The PF results are used as reference as they represent the ``real" system state.

\begin{figure*}[t]
\begin{tabular}{cc}
  \includegraphics[width=0.49\textwidth]{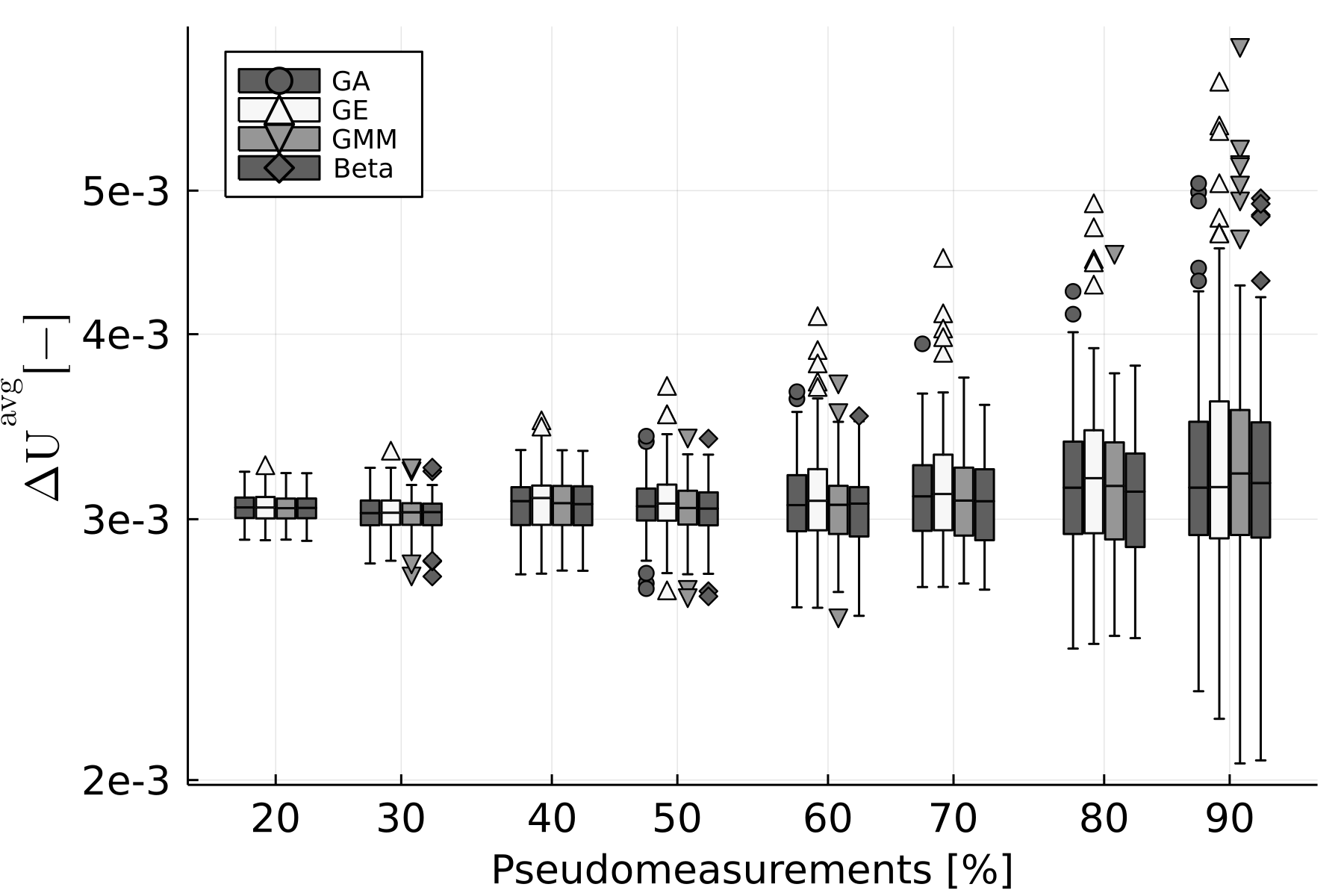}\label{fig:c1_avgvolt} &   \includegraphics[width=0.49\textwidth]{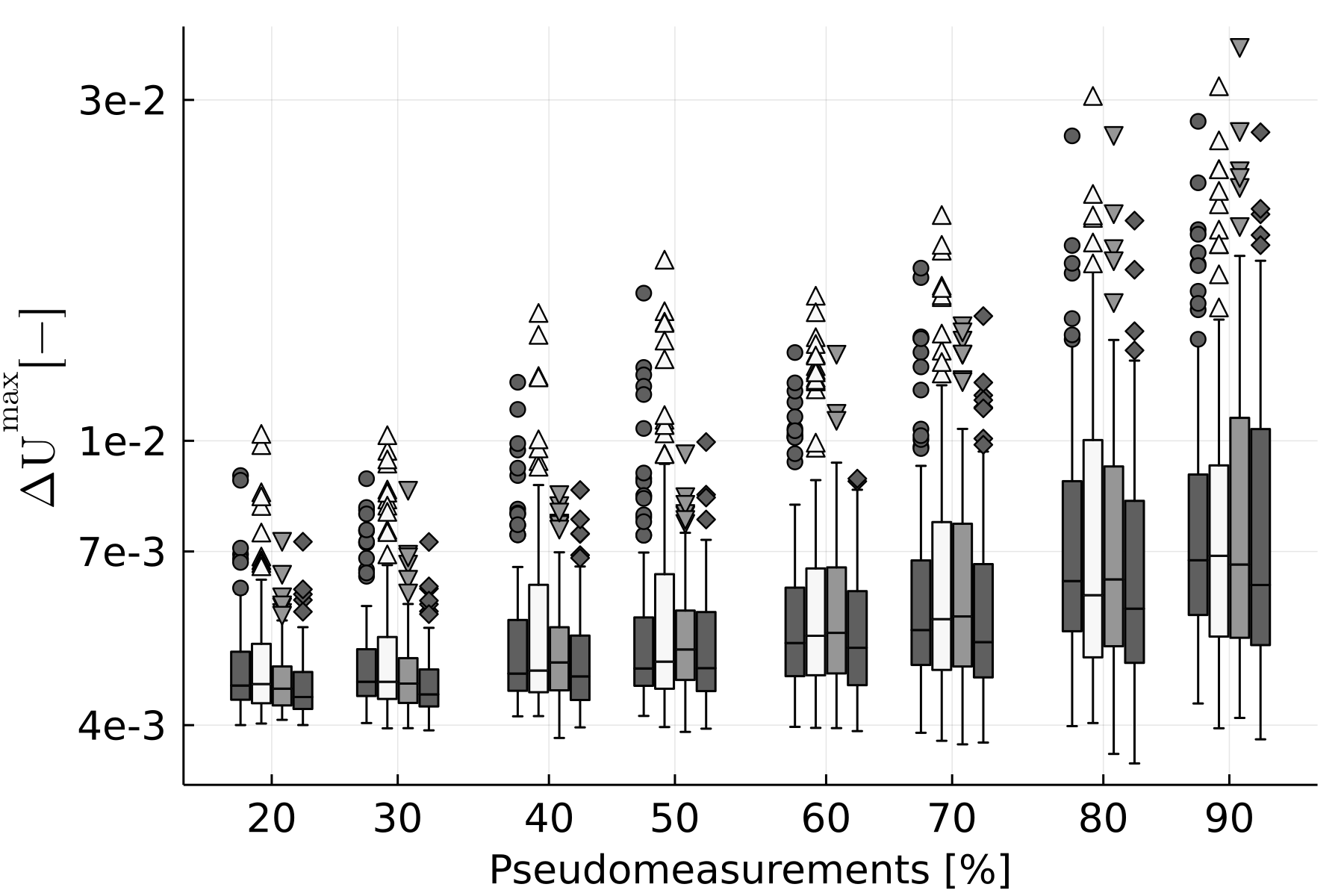}\label{fig:c1_maxvolt} \\
(a)  & (b)  \\
\includegraphics[width=0.49\textwidth]{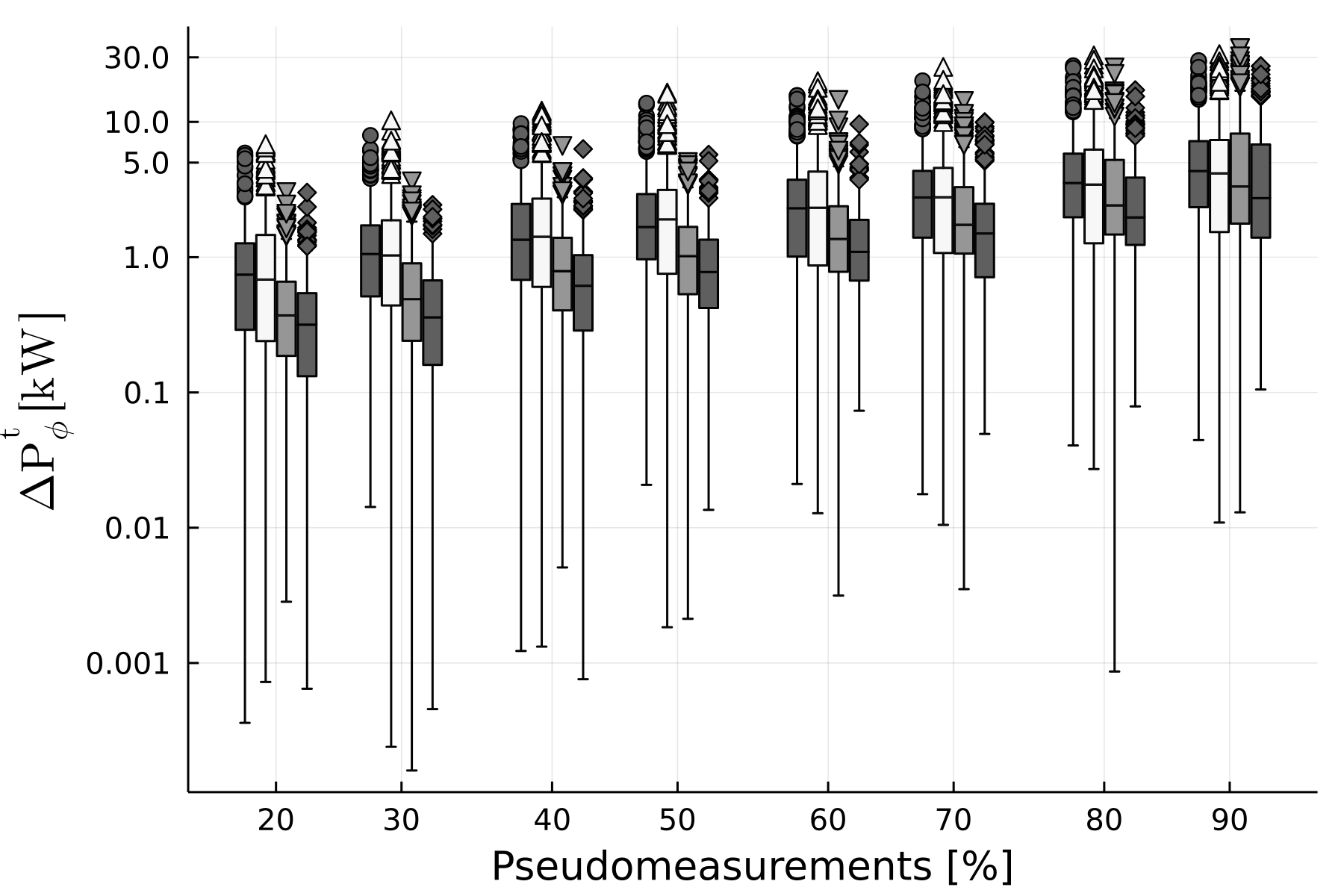}\label{fig:c1_power} & \includegraphics[width=0.49\textwidth]{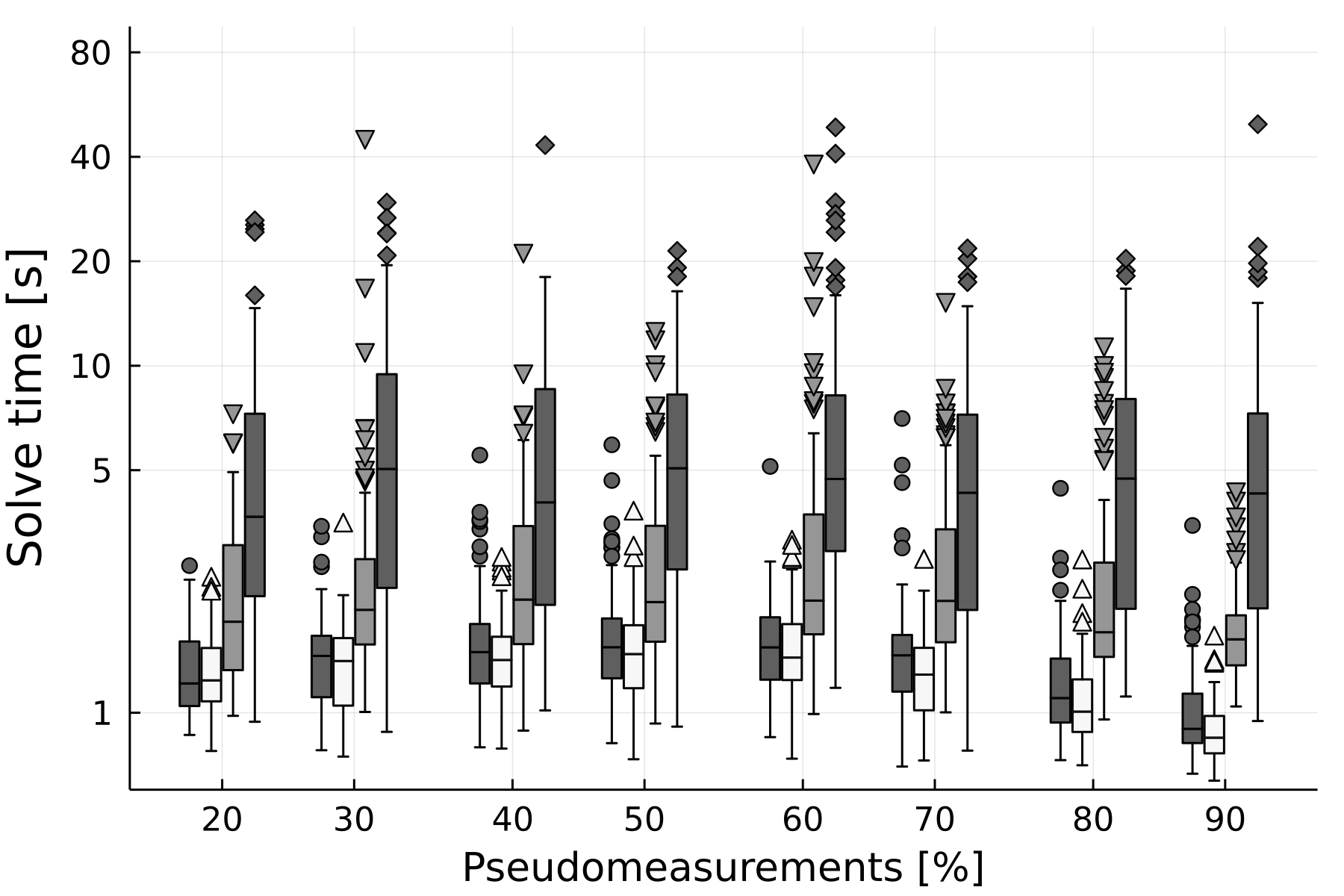}\label{fig:c1_time}\\
(c) & (d)  \\
\end{tabular}
\caption{Voltage magnitude difference $\Delta U^{\text{avg}}$ (a)- $\Delta U^{\text{max}}$(b), in p.u., and power difference $\Delta P^{\text{t}}_{\phi}$ (c) between SE and PF results; SE solve times (d) - Case Study I: Beta distr., Q independent of P. Figures are in logarithmic scale. }
\label{fig:cs_1}
\end{figure*}

As such, there are three $\Delta P^{\text{t}}$ values and one $\Delta U^{\text{avg}}$ and $\Delta U^{\text{max}}$ for each PF/SE calculation. Simulations are performed for increasing ratios of users with pseudo-measurements over users with SM. For each ratio value, 100 SE calculations are performed; each of them is  characterized by a different random sample of users to which pseudo-measurements are assigned, and different random power samples from the pdf in the PF stage, which are then used to generate the SE input. Results are shown for ratios in the range of 20\%-90\%: the differences between the uncertainty models for lower values is negligible, while the case with 100\% pseudo-measurements would lead to an underdetermined problem, as there is no reference voltage value/measurement, and has been omitted.
Table \ref{tab:case_studies_ngse} summarizes the settings in the different case studies.

\begin{table}[b]
\centering
\caption{Features of the different case studies.}
\begin{tabular}{l|c|c}
 & Distribution & Constant power factor \\ \hline
 Case study I & $Beta$  &   No  \\ 
 Case study II & $Beta$  &  Yes \\ 
 Case study III  & $Polynomial$  & Yes \\  

\end{tabular}\label{tab:case_studies_ngse}
\end{table}

The nonlinear solver used for the calculations is the open-source Ipopt v3.12.10 \cite{ipopt}, with its HSL MA27 subroutine \cite{HSL}. Simulations are run on a 64-bit machine with Intel(R) Xeon(R) CPU E5-4610 v4 @1.80GHz, 32 GB RAM, using Julia 1.6 and PowerModelsDistributionStateEstimation.jl 0.6.0.

\subsection{Case Study I and II}

\begin{figure*}[t!]
\begin{tabular}{cc}
  \includegraphics[width=0.49\textwidth]{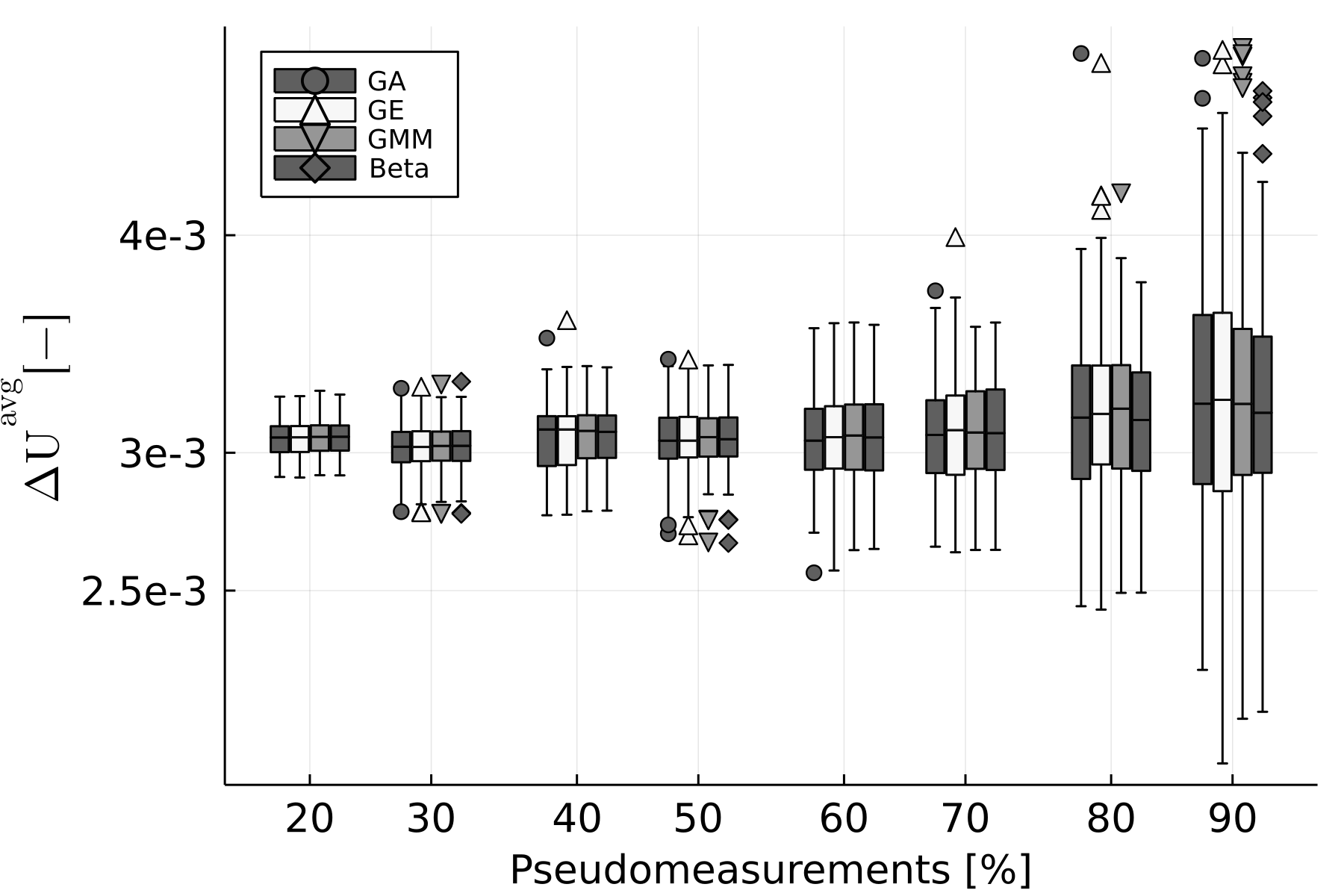}\label{fig:c2_avgvolt} &   \includegraphics[width=0.49\textwidth]{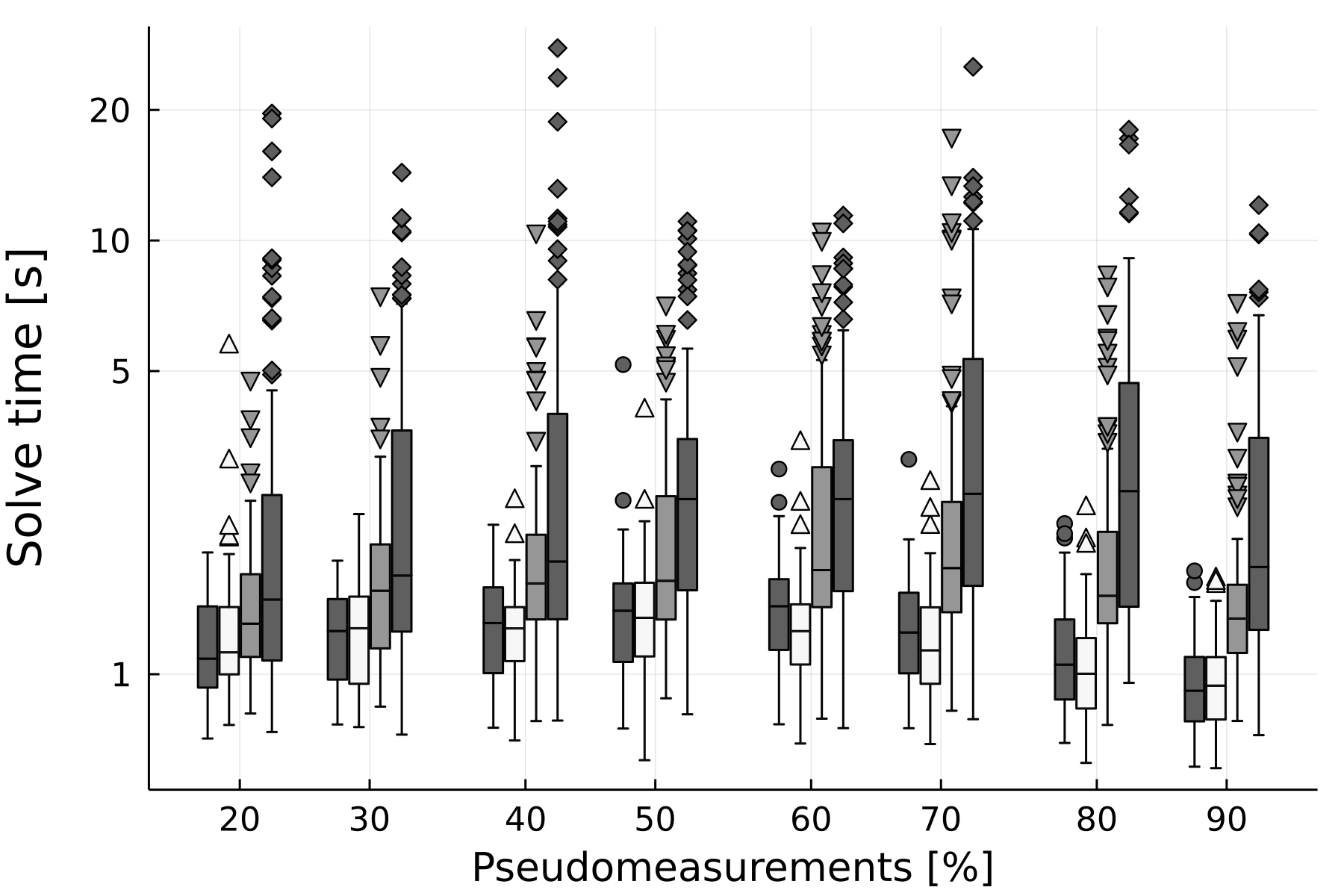}\label{fig:c2_time} \\
\end{tabular}
\caption{Avg. voltage magnitude difference $\Delta U^{\text{avg}}$ and SE solve times - Case Study II: Beta distr., constant power factor. Figures are in log. scale. }
\label{fig:cs_2}
\end{figure*}

\begin{figure*}[b]
\begin{tabular}{cc}
  \includegraphics[width=0.49\textwidth]{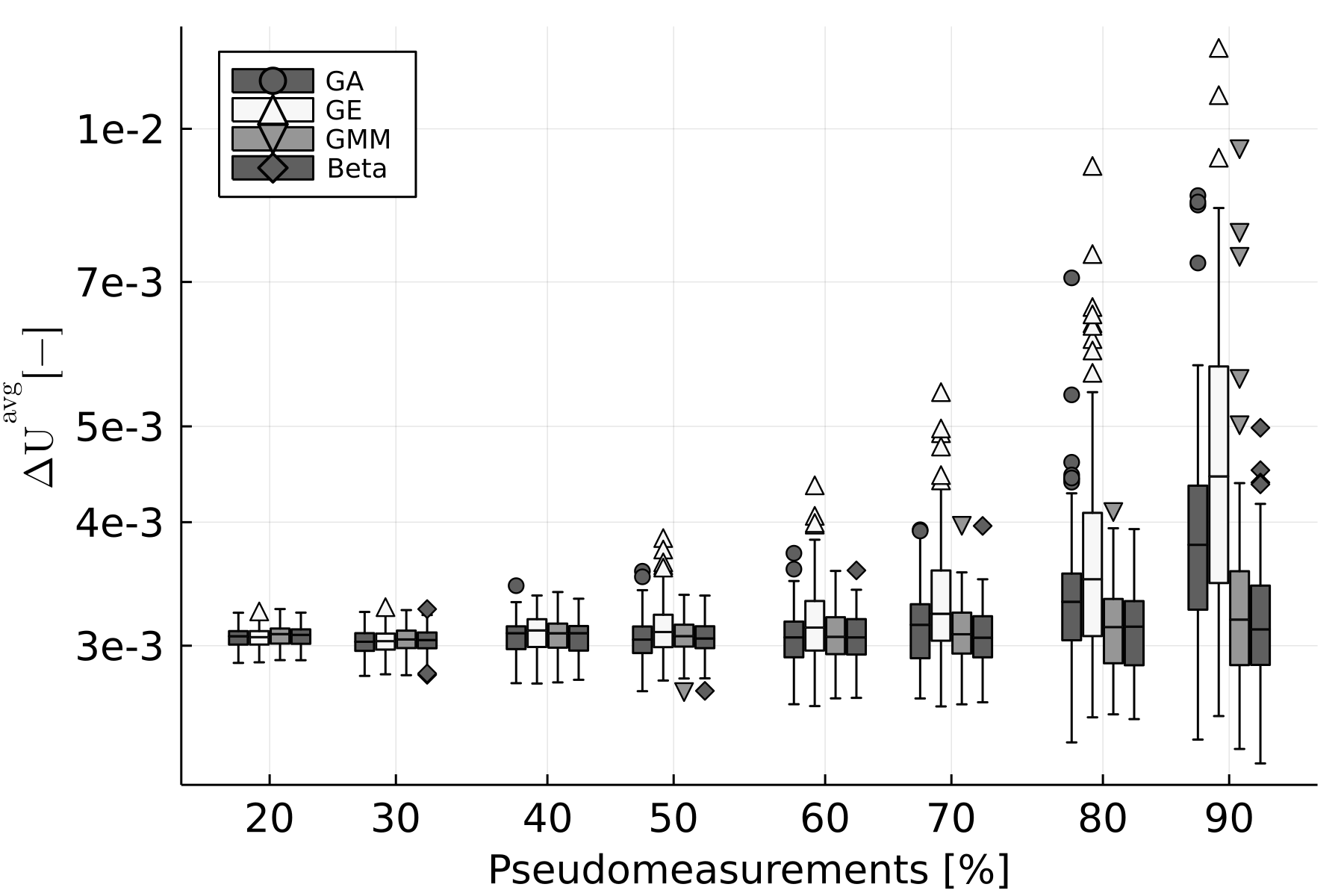}\label{fig:poly_avgvolt} &   \includegraphics[width=0.49\textwidth]{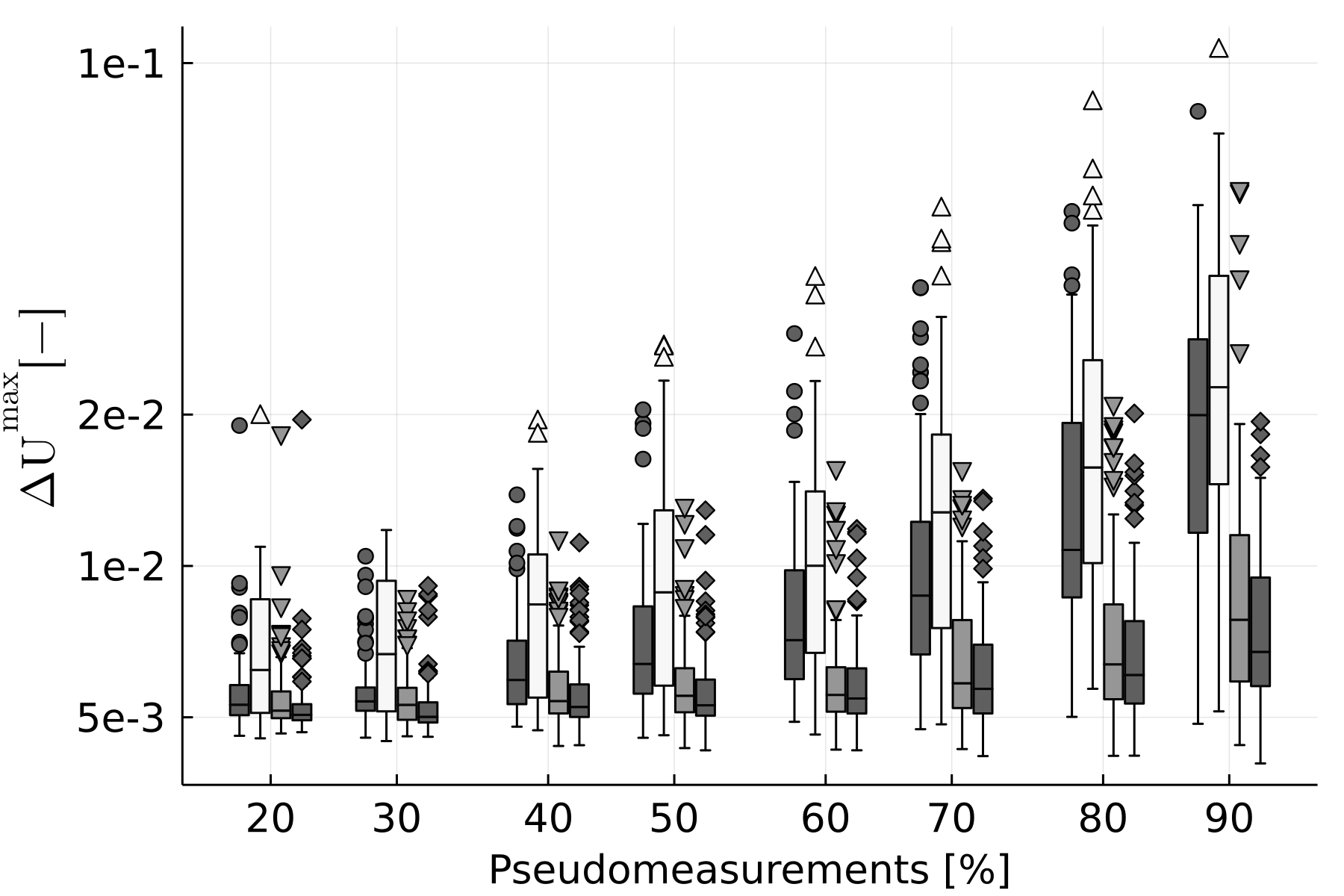}\label{fig:poly_maxvolt} \\
(a)  & (b)  \\
\includegraphics[width=0.49\textwidth]{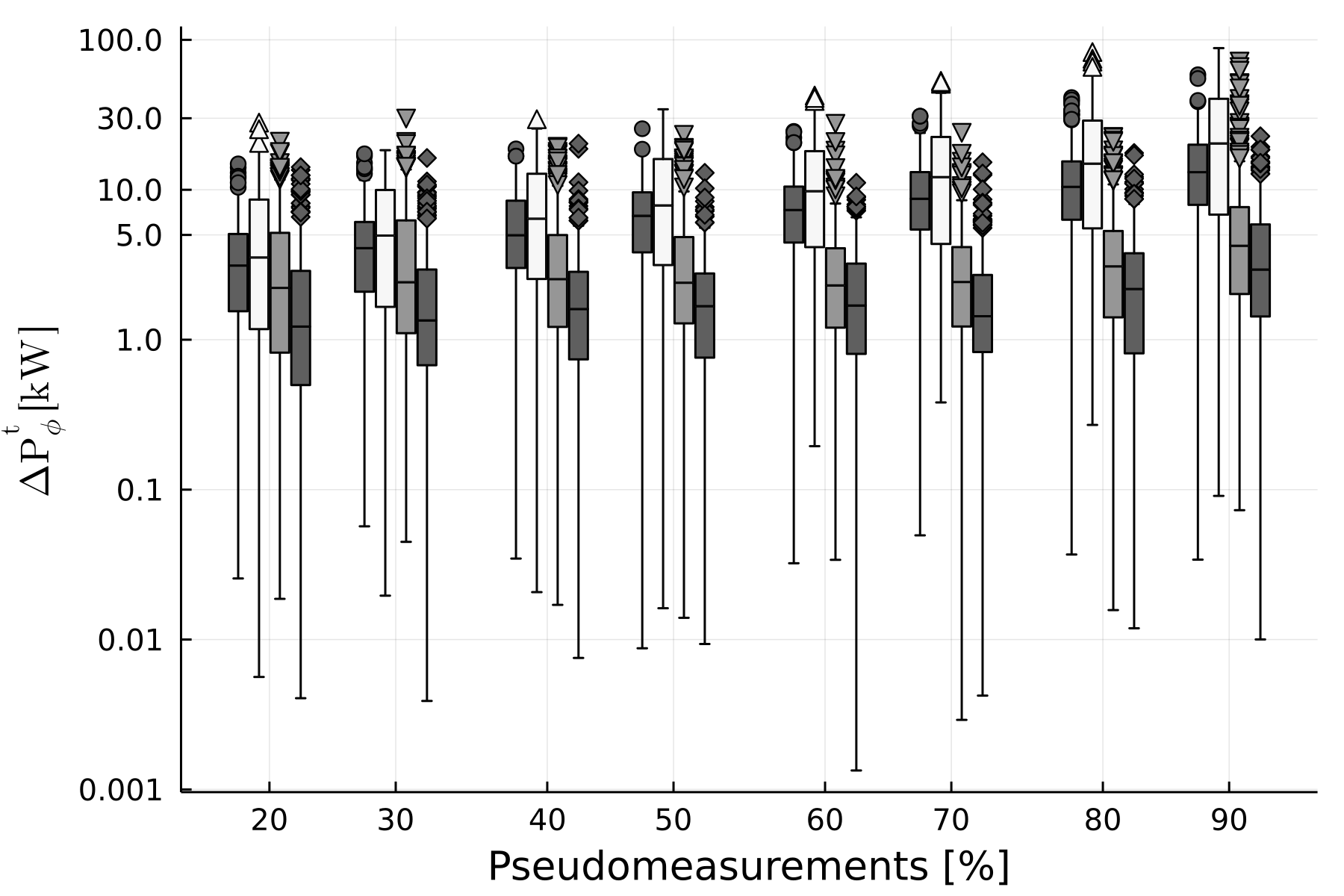}\label{fig:poly_power} & \includegraphics[width=0.49\textwidth]{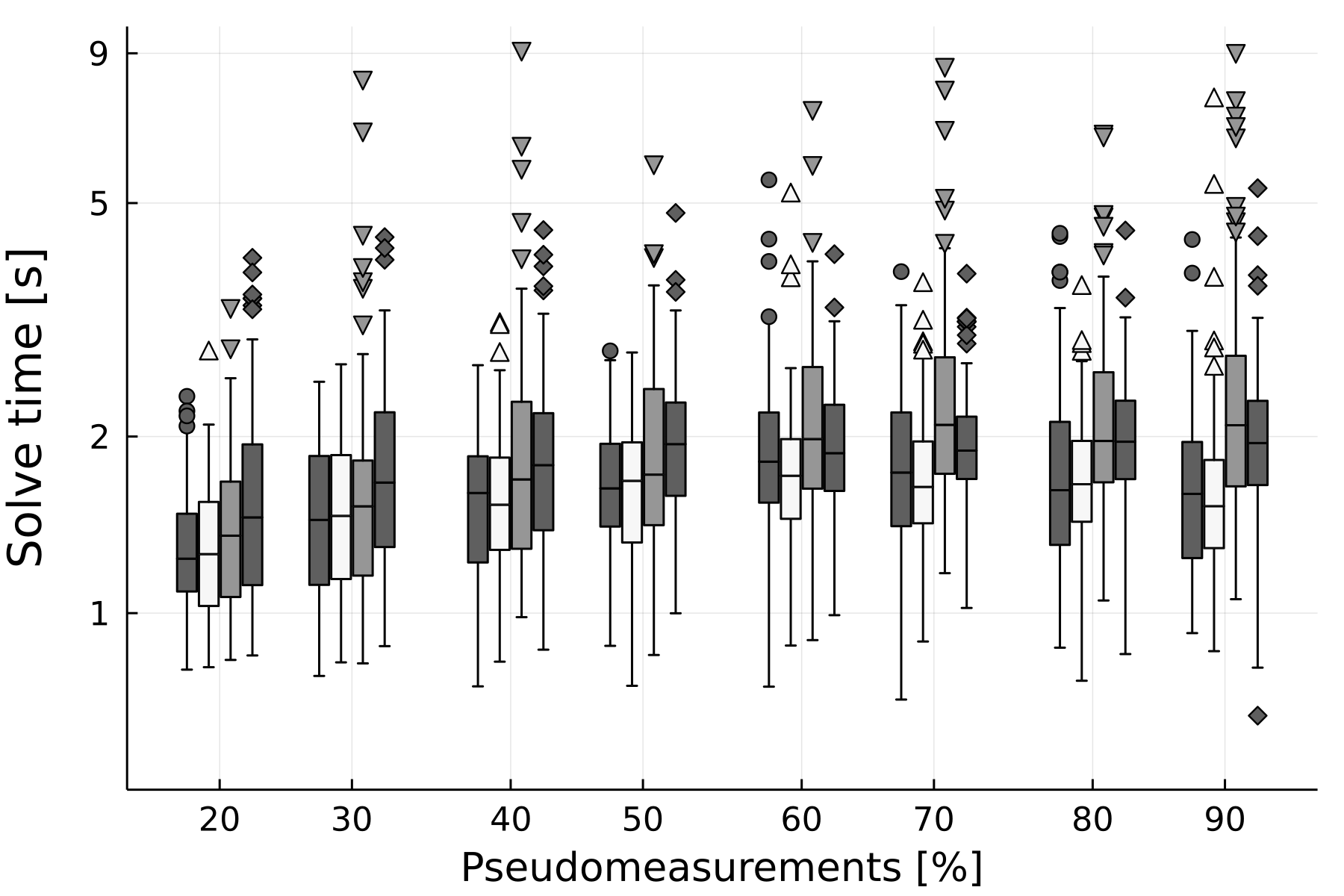}\label{fig:poly_time}\\
(c) & (d)  \\
\end{tabular}
\caption{Voltage magnitude difference $\Delta U^{\text{avg}}$ (a)- $\Delta U^{\text{max}}$(b), in p.u., and power difference $\Delta P^{\text{t}}_{\phi}$ (c) between SE and PF results; SE solve times (d) - Case Study III: Polynomial distr., constant power factor. Figures are in logarithmic scale.}
\label{fig:cs_3}
\end{figure*}

Fig.~\ref{fig:cs_1}-\ref{fig:cs_2} show the results for case study I and II, respectively. Voltage and power results of case II are very similar to those of case I. As such, for the sake of the page limit, they are only partially shown in Fig.~\ref{fig:cs_2}. Fig.~\ref{fig:cs_1} shows that the differences between SE and PF results increase for increasing amounts of pseudo-measurements, which is expected given their lower accuracy with respect to real measurements. The SE is quite accurate: as an indication, the voltage reference is 230~V, so a difference of 0.005~p.u. corresponds to 1.15~V, which is rarely exceeded, except the cases with very high amounts of pseudo-measurements. The Beta distribution leads to the best results, followed by the GMM and then the Gaussian models, as expected. However, the differences are minimal, with the approximation only displaying a slight tendency to have more outliers, i.e., occasionally, they yield relatively large errors.  

Solve times do not increase significantly for increasing percentages of pseudo-measurements, and while the Beta solve times are noticeably longer than for the other models, they almost never exceed 40 seconds in case I and are even lower in case II. The latter is due to the constant pf constraint. These solve times seem acceptable for static DSSE, where the SM measurement collection rate is hardly higher than 15 minutes. As such, even though the accuracy improvements are marginal, nothing seems to prevent the adoption of the MLE-SE. Furthermore, the shape of the chosen Beta pdf is quite similar to that of its Gaussian approximation. This does not hold with the Polynomial distribution, and the results of the next case study differ significantly.

\subsection{Case Study III}

Fig.~\ref{fig:cs_3} shows the results for Case Study III. In this case, the MLE-SE is much more accurate than the WLS-SE, even at relatively low pseudo-measurement ratios. This is attributable to the shape of the Polynomial distribution, which is strongly non-Gaussian. The solve times of the MLE-SE and the WLS-SE are comparable and are only a few seconds. As such, the MLE-SE should be the estimator of choice in this case.

\subsection{Summary of the Results}

To summarize, the MLE-SE is slower than the WLS-SE, as expected, but its computational time appears acceptable, especially as the measurement refresh rate in the DS is typically low. Whether the accuracy improvement is significant enough to justify the extra computational effort depends on the practical scenario and SE requirements. In the examined cases, larger accuracy gains occur in the Polynomial case with constant power factor. Factors that influence the gains in accuracy are (mainly) the amount of deviation of the original uncertainty from a Gaussian one, and the absolute power value of the samples. Since constant power factor is a common inverter setting, this appears a good practical application for rooftop PV systems. 
Finally, it is clear that both MLE-SE and WLS-SE work better when the provided pdfs are correct and known in advance. Discrepancies between the real and the provided pdfs will in general lead to the deterioration of the SE accuracy. Nevertheless, the better the pdf approximates the original distribution, the better the SE results. This can be noted, for example, by comparing the GE and GA results to the GMM's. Obtaining realistic pdfs seems feasible, especially when forecasts are updated over time, and for measurement devices for which the error distribution is known. 

\section{Conclusions}\label{sec:conclusions}
A novel method was presented, that allows to exactly model non-Gaussian (pseudo-)measurement uncertainties for state estimation in unbalanced distribution networks. The method is based on maximum-likelihood estimation, and an implementation is made available as part of an open-source tool. The ``optimization-first" state estimation formulation allows to add constraints that particularly suit the modelling of non-monitored PV injection. Case studies illustrate the trade-off between the increased computational effort and improved accuracy that stem from the adoption of such method, with respect to making standard Gaussian assumptions. The case studies show that the maximum-likelihood state estimation is indeed more accurate than the standard Gaussian WLS state estimation. The accuracy increase is scenario-dependent. As such, the flexibility of the method/tool is beneficial: it allows to easily assess the trade-off for multiple scenarios, providing decision support to find suitable uncertainty models. Finally, the solve times seem acceptable for practical DSSE applications in all the examined cases, which makes the proposed method suitable for operations.

\section*{Acknowledgment}
The authors would like to thank Bruno Macharis and Andy Gouwy from Fluvius and Arpan Koirala from KU Leuven for the valuable discussions and contributions.

\bibliographystyle{IEEEtran}
\bibliography{IEEEabrv,Bibliography.bib}

\end{document}